\documentclass[10pt]{article}
\usepackage{amsmath,amssymb,graphicx,paralist}

\newcommand{\comment}[1]{}

\begin{document}

\title{Can Modified Gravity (MOG) explain the speeding Bullet (Cluster)?}

\author{J. W. Moffat$^{1,2}$ and V. T. Toth$^1$\\
$^1$Perimeter Institute for Theoretical Physics,\\Waterloo, Ontario N2L 2Y5, Canada\\
$^2$Department of Physics, University of Waterloo,\\Waterloo, Ontario N2L 3G1, Canada}

\maketitle

\begin{abstract}
We apply our scalar-tensor-vector (STVG) modified gravity theory (MOG) to calculate the infall velocities of the two clusters constituting the Bullet Cluster 1E0657-06. In the absence of an applicable two-body solution to the MOG field equations, we adopt an approximate acceleration formula based on the spherically symmetric, static, vacuum solution of the theory in the presence of a point source. We find that this formula predicts an infall velocity of the two clusters that is consistent with estimates based on hydrodynamic simulations.
\end{abstract}

Hydrodynamic simulations of the colliding galaxy clusters 1E0657-06 (known as the Bullet Cluster) indicate that at a distance of 4.6~Mpc between the two clusters, an infall velocity of 3,000~km/s is required \cite{2008RvMA...20..228M} in order to achieve the observed X-ray brightness and morphology of the cluster. Recently, it has been argued \cite{2010arXiv1003.0939L} that such a high infall velocity is incompatible with predictions of the standard $\Lambda$CDM model of cosmology.

Our scalar-vector-tensor (STVG \cite{Moffat2006a}) modified gravity theory (MOG), which predicts significant deviations from Newtonian gravity at mass scales of $10^6--10^7~M_\odot$ and beyond, can account for the apparent discrepancy between the gravitational lensing and X-ray maps of the colliding clusters \cite{Brownstein2007} without requiring collisionless dark matter. In light of this encouraging result, it is natural to ask if MOG is perhaps also capable of predicting the correct collision velocity.

The answer to this question may seem straightforward: since MOG predicts stronger gravity on intergalactic scales, a greater collision velocity should result. However, due to the absence of dark matter, the magnitudes of the gravitating masses are reduced. Symbolically speaking, even as we increase the effective gravitational constant $G_\mathrm{eff}\gg G_N$ (where $G_N$ is the Newtonian value), the absence of dark matter may very well mean that $G_\mathrm{eff}M_\mathrm{baryon}\simeq G_N(M_\mathrm{baryon}+M_\mathrm{CDM})$ (where $M_\mathrm{baryon}$ and $M_\mathrm{CDM}$ denote the baryonic and cold dark matter masses, respectively) and the overall gravitational accelerations remain approximately the same.

Nonetheless, it is possible for MOG to account for the observed infall velocities, for two reasons. First, at extragalactic distance scales, for very large source masses, the effective MOG gravitational constant may be significantly larger than the presumed ratio between cold dark matter and baryonic matter. Second, a nonlinear interaction between two large masses may further enhance the strength of gravity, producing an additional contribution to acceleration.

Unfortunately, any further analysis is hampered by the fact that to date, the only reliable solution to the MOG field equations is the spherically symmetric, static vacuum solution. Although this solution yielded valuable results in cases that are well approximated by a model with a single source orbited by massless test particles (e.g., a spiral galaxy dominated by the central bulge) and it was even extended to the case of small perturbations in an homogeneous background, none of this is directly applicable to the Bullet Cluster, which, even in a crude approximation, is a genuine two-body problem. For such problems, not even an approximate solution exists presently in the context of MOG. (Note, however, that such a solution is in development \cite{Moffat2010a}.)

Despite these difficulties, the importance of the Bullet Cluster warrants a somewhat speculative analysis, in which the spherically symmetric, static vacuum solution of the MOG field equations is used to obtain at least a rough understanding of the dynamics of the infall of these two large clusters.

The calculation must begin by estimating the amount of mass present in the two clusters prior to their collision. In addition to the observed post-collision cluster masses, this must also include mass that may have been segregated from the clusters or otherwise dispersed by the collision. One must also ensure that the correct dynamical masses (which include the presumed dark matter content) are used for Newtonian gravity calculations, but for modified gravity with no dark matter, only the baryonic mass content must be used.

Mass estimates for the main cluster, the ``bullet'', and surrounding matter were provided in \cite{Bradac:2006er}: $M_\mathrm{main}=(2.8\pm 0.2)\times 10^{14}~M_\odot$ and $M_\mathrm{bullet}=(2.3\pm 0.2)\times 10^{14}~M_\odot$, respectively. These dynamical masses were inferred from lensing observations. The ratio of baryonic mass to dynamical mass within the observational instrument's field of view was estimated as $M_\mathrm{baryon}/M_\mathrm{total}=0.14\pm 0.03$.

Given these masses, it is easy to calculate the Newtonian infall velocities of the two clusters at a given distance $r$. Assuming that the clusters were at relative rest at infinity, the velocity is just the escape velocity of the two clusters in question. The gravitational potential energy at $r$ is $-G_NM_\mathrm{main}M_\mathrm{bullet}/r$, and this must be equal to $-1$ times the combined kinetic energy, $(1/2)(M_\mathrm{main}v_\mathrm{main}^2+M_\mathrm{bullet}v_\mathrm{bullet}^2)$. On the other hand, the combined momenta of the two clusters must be zero: $M_\mathrm{main}v_\mathrm{main}+M_\mathrm{bullet}v_\mathrm{bullet}=0$. Solving these two equations, we obtain an infall velocity of $v_\mathrm{main}-v_\mathrm{bullet}=980\pm 40$~km/s at a distance of 4.6~Mpc. This is over 2,000~km/s short of the velocity required to produce the recorded X-ray observations.

To perform the same calculations in a modified gravity theory without cold dark matter, one needs to use mass estimates that do not include dark matter. For this purpose, we use the estimated $M_\mathrm{baryon}/M_\mathrm{total}$ ratios to obtain the revised cluster masses of $M_\mathrm{main}=(3.9\pm 0.3)\times 10^{13}~M_\odot$ and $M_\mathrm{bullet}=(3.2\pm 0.3)\times 10^{13}~M_\odot$.

The MOG acceleration law for the acceleration $a$ of a massless test particle at distance $r$ from a point source with mass $M$ is written as
\begin{equation}
a=\frac{G_NM}{r^2}\left[1+\alpha-\alpha\left(1+\frac{r}{r_0}\right)e^{-r/r_0}\right],
\end{equation}
with $\alpha$ and $r_0$ given by \cite{Moffat2007e}:
\begin{align}
\alpha\simeq&\frac{19M}{(\sqrt{M}+25000)^2}\\
r_0\simeq&\frac{\sqrt{M}}{6250}~\mathrm{kpc},
\end{align}
and $M$ given in units of $M_\odot$. This acceleration law combines a Newtonian acceleration with increased strength ($G\rightarrow(1+\alpha)G_N$) and a {\em repulsive} force with a finite range $r_0$, due to the presence of a massive vector field in the theory.

Using this acceleration law and treating each cluster as a point mass moving in the MOG gravitational field of the other cluster, we obtain an infall velocity of $\sim 1620$~km/s. Although this is significantly larger than the Newtonian estimate (despite the fact that only the baryonic masses of the clusters were used in the solution), it is still not sufficient to account for the X-ray observations.

However, it is naive to assume that a linear superposition of two point source solutions will correctly approximate the solution of the two-body problem in MOG. MOG predicts an increase in the coupling strength between mass and geometry for a massive source; it is likely, then, that a similar increase in the coupling between geometry and a passive gravitational mass must be present.

In the absence of even an approximate solution of the MOG field equations in two-body case, we must resort to a somewhat {\em ad-hoc} modification of the MOG acceleration law when two heavy bodies are considered. For a single source, the parameter $\alpha$ determines the effective gravitational constant at a large distance from the source: $\lim\limits_{r\rightarrow\infty}a=(1+\alpha)G_NM/r^2$. For two heavy bodies characterized individually by $\alpha_1$ and $\alpha_2$, we adopt the formula $\alpha=(1+\alpha_1)(1+\alpha_2)-1$. We assume a similar increase in the range of the vector force associated with each source: $r_{0(1)}\rightarrow r_{0(1)}(1+\alpha_2)$ and $r_{0(2)}\rightarrow r_{0(2)}(1+\alpha_1)$.

Using this modified MOG acceleration formula in the case of the Bullet Cluster, we obtain an infall velocity of $\sim 3,300$~km/s when the two clusters are 4.6~Mpc apart, assuming that they were at relative rest at infinity.

It must be strongly emphasized that this result is qualitative at best: it does not represent a proper solution of the MOG field equations in the presence of two heavy sources. Nonetheless, the fact that the calculated infall velocity is of the same magnitude as that which is required to reproduce the clusters' observed hydrodynamic behavior is encouraging, and offers strong motivation to continue our research in this direction.


The research was partially supported by National Research Council of Canada. Research at the Perimeter Institute for Theoretical Physics is supported by the Government of Canada through NSERC and by the Province of Ontario through the Ministry of Research and Innovation (MRI).

\bibliography{refs}
\bibliographystyle{unsrt}

\end{document}